# Comments on Superluminal Laser Pulse Propagation

P. Sprangle, J.R. Peñano[1], and B. Hafizi[2]

*Naval Research Laboratory, Plasma Physics Division, Washington D.C.*

**Researchers claim to have observed superluminal (faster than light) propagation of a laser pulse in a gain medium by a new mechanism in which there is no distortion of the pulse. Our analysis shows that the observed mechanism is due to pulse distortion arising from a differential gain effect and should not be viewed as superluminal propagation.**

In a recent article in *Nature* [1], titled ***Gain-assisted superluminal light propagation***, experimentalists reported observing a laser pulse propagating faster than the speed of light through a gas cell which served as a amplifying medium. They state that the "peak of the pulse appears to leave the cell before entering it". This article generated a great deal of press attention around the world. For example, reports appeared in *CNN* and in newspapers such as the *Los Angeles Times, New York Times, Washington Post, South China Morning Post, India Today,* and the *Guardian of London* [2].

The objective of this paper is to analyze, discuss and comment on the conclusions reached in Ref. 1 and related studies [3]. We show that the theory on which this superluminal interpretation was based is inaccurate because of an inconsistency in the ordering of the terms kept in the analysis. Our analysis of the experiment shows that the mechanism responsible for the so-called "superluminal propagation" can be attributed to *differential gain* of the laser pulse. That is, the front of the laser pulse is amplified more than the back, causing the pulse shape to be tilted towards the front. The leading edge of the distorted pulse, however, propagates at the speed of light, as one would expect. The authors of Ref. 1 specifically discount this explanation. They claim that superluminal propagation is observed "while the shape of the pulse is preserved" and "the argument that the probe pulse is advanced by amplification of its front edge does not apply" in the present experiment. Our analysis indicates that these claims are incorrect.

[1]LET Corp., 4431 MacArthur Blvd., Washington, DC 20007

[2]Icarus Research, Inc., P.O. Box 30780, Bethesda, MD 20824-0780

In the experiment a long laser pulse was passed through an amplifying medium consisting of a specially prepared caesium gas cell of length $L = 6$ cm, as depicted in Fig. 1. The laser pulse of duration $T = 3.7$ μsec was much longer (1.1 km) than the gas cell, so that at any given instant only a small portion of the pulse was inside the cell. By measuring the pulse amplitude at the exit, they claim that both the front and the back edges of the pulse were shifted forward in time by the same amount relative to a pulse that propagated through vacuum.

The following analysis considers a laser pulse propagating in a general dispersive and amplifying medium characterized by a frequency dependent complex refractive index $n(\omega)$. To determine the evolution of the pulse envelope we represent the laser electric field amplitude as

$$E(z,t) = (1/2)\ A(z,t)\exp\big(i(k_o\, z - \omega_o\, t)\big) + c.c., \tag{1}$$

where $A(z,t)$ is complex and denotes the slowly varying laser pulse envelope, $k_o = \omega_o\, n(\omega_o)/c$ is the complex wavenumber, $\omega_o$ is the carrier frequency and c.c denotes the complex conjugate. The field is polarized in the transverse direction and propagates in the z-direction. Since $k_o$ is complex, the factor $\exp\big(-\mathrm{Im}(k_o)\, z\big)$ represents an overall amplification of the pulse at frequency $\omega_o$ which does not result in pulse distortion. The deviation of the refractive index from unity, i.e., $\Delta n(\omega) = n(\omega) - 1$, is small in the reported experiment and it is therefore legitimate to neglect the reflection of the laser pulse at the entrance and exit of the gas cell.

The envelope equation describing the evolution of a laser pulse propagating in a general dispersive medium has been derived elsewhere [4]. For a one-dimensional laser pulse the envelope equation is given by

$$\left(\frac{\partial}{\partial z} + \frac{1}{c}\frac{\partial}{\partial t}\right) A(z,t) = -\left(\kappa_1 \frac{\partial}{\partial t} + \frac{i}{2}\kappa_2 \frac{\partial^2}{\partial t^2} + \ldots\right) A(z,t), \tag{2}$$

where $\kappa_\ell = \left(\partial^\ell \kappa(\omega) / \partial \omega^\ell\right)_{\omega=\omega_o}$, $\ell = 1,2,\ldots$, $\kappa(\omega) = \omega\,\Delta n(\omega)/c$ and the laser pulse envelope at the input to the amplifying medium $A(z=0,t)$ is assumed given. Equation (2) is derived by substituting the field representation given by Eq. (1) into the linearized wave equation and performing a spectral analysis [4,5] that involves expanding the refractive index about the carrier frequency $\omega_o$ and neglecting reflected waves. These assumptions are valid when the refractive index is close to unity and the spectral width of the pulse is narrow. We limit our analysis to terms of order $\kappa_2$, i.e., to lowest order in group velocity dispersion; this is sufficient for the present purpose. In a vacuum, $\Delta n(\omega) = 0$ so that the right hand side of Eq. (2) vanishes and the envelope is given by $A(z,t) = A(0, t - z/c)$, indicating that the pulse propagates without distortion with velocity c.

To solve for the pulse envelope in a general dispersive medium we Fourier transform Eq. (2) in time and solve the resulting differential equation in z for the transformed envelope. Inverting the transformed envelope yields the solution

$$A(z,t) = \frac{1}{\sqrt{2\pi}} \int_{-\infty}^{\infty} d\nu\, \hat{A}(0,\nu) \exp(-i\nu(t - z/c)) \exp\left(i\kappa_1 \nu\, z + i\kappa_2 \nu^2 z/2\right), \quad (3)$$

where $\hat{A}(0,\nu)$ is the Fourier transform of the envelope at z = 0, and $\nu$ is the transform variable.

In the experiment the values of the parameters are such that the following inequalities hold: $1 >> |\kappa_1 \nu z| >> |\kappa_2 \nu^2 z/2|$, where $\nu \approx 1/T$ and $z = L$. To correctly evaluate the integral in Eq. (3), the exponentials in the small quantities should be expanded to an order of approximation consistent with Eq. (2), otherwise unphysical solutions may result. For example, if the higher order term $\kappa_2 \nu^2 z/2$ is neglected in Eq. (3), the laser envelope is given by

$$A(z,t) = \frac{1}{\sqrt{2\pi}} \int_{-\infty}^{\infty} d\nu\, \hat{A}(0,\nu) \exp(-i\nu(t - z/c)) \exp(i\kappa_1 \nu\, z). \quad (4)$$

Equation (4) can be integrated exactly to give

$$A(z,t) = A(0, t - z/v_p)\,. \quad (5)$$

The quantity $v_p = [\partial(\omega n/c)/\partial\omega]^{-1} = c/(1 + c\kappa_1)$ defines the group velocity of a pulse in a dispersive medium. However, in addition to cases where $v_p$ is abnormal i.e., greater than *c* or negative, there are other instances in which the concept of group velocity does not represent the pulse velocity. These include situations where the interaction length, L, is less than the phase mixing length associated with the spectral components of the pulse, or when L is much less than the pulse length. These conditions apply in the experiment of Ref. 1.

The exact solution, given by Eq. (5), to the approximate envelope equation can lead to unphysical conclusions since it implies that the pulse propagates undistorted with velocity $v_p$ [3]. For example, if $-1 < c\kappa_1 < 0$, the pulse velocity exceeds *c*. For the parameters in Ref. 1, however, $c\kappa_1$ is essentially real and < -1, giving a negative pulse velocity, $v_p = -c/310$. They have, as well as others, ascribed physical meaning to this by considering the delay time $\Delta T = L/v_p - L/c$, defined as the difference in the pulse transit time in the gain medium and in vacuum [1,6]. Since $v_p = -c/310$ in Ref. 1, the delay time is negative, implying superluminal propagation. *This error is due to retaining terms beyond the order of the approximation*. That is, the exponential factor in Eq. (4) contains terms to all orders in $\kappa_1 \nu z$ while terms of order $\kappa_2 \nu^2 z$ and higher are neglected. For example, this is equivalent to keeping terms proportional to $(\kappa_1 \nu z)^2$ while neglecting terms proportional to $\kappa_2 \nu^2 z$ which are of the same order.

To avoid this incorrect conclusion it is necessary to solve Eq. (3) by keeping the order of approximation consistent. Expanding the exponential terms in Eq. (3) to second order gives

$$A(z,t) = (1/\sqrt{2\pi}) \int_{-\infty}^{\infty} d\nu\, \hat{A}(0,\nu) \left(1 + i\kappa_1 z\nu + (1/2)(i\kappa_2 z - \kappa_1^2 z^2)\nu^2\right)$$

$$\times \exp\left(-i\nu(t - z/c)\right). \qquad (6)$$

Equation (6) can be integrated to give

$$A(z,t) = \left(1 - \kappa_1 z \frac{\partial}{\partial t} - \frac{1}{2}\left(i\kappa_2 z - \kappa_1^2 z^2\right)\frac{\partial^2}{\partial t^2} + \ldots\right) A(0, t - z/c). \qquad (7)$$

In Eq. (7) the first term on the right hand side denotes the vacuum solution, the second term represents lowest order differential gain, while the third and higher order terms are small and denote higher order effects. The result in Eq. (7) shows that the pulse propagates at the speed of light while undergoing differential gain. The quantity $\kappa_1$ can be negative in the presence of gain or absorption. In the case of gain, when $\kappa_1 < 0$, the front portion of the pulse is amplified more than the back. Note that the differential gain effect, i.e., the first order $\partial / \partial t$ term in Eq. (7), can be recovered from Eq. (5) through a Taylor expansion. However, this is simply equivalent to expanding Eq. (5) so that the proper order of approximation is recovered, as was done in deriving Eq. (7).

The results of this analysis may be used to interpret the experiment. The susceptibility of the medium used in Ref. 1 has the following form near the resonance frequencies

$$\chi(f) \cong \frac{\Delta n(\omega)}{2\pi} = \frac{M_1}{f - f_1 + i\gamma} + \frac{M_2}{f - f_2 + i\gamma}, \tag{8}$$

where $M_{1,2} > 0$ are related to the gain coefficients. The susceptibility in Eq. (8) represents a medium with two gain lines of spectral width $\gamma$ at resonance frequencies $f_1$ and $f_2$. The gain spectrum for $M_{1,2} = M = 0.18$ Hz, $f_1 = 3.5 \times 10^{14}$ Hz, $f_2 = f_1 + 2.7$MHz and $\gamma =$ 0.46 MHz, is shown in Fig. 2(a) (solid curve). For these parameters the deviation of the refractive index from unity $\Delta n(\omega)$ shown in Fig. 2(b) closely approximates that in Fig. 3 of Ref. 1. The input laser pulse envelope, which approximates the experimental pulse, is taken to have the form

$$A(z=0,t) = \begin{cases} a_o \sin^2(\pi t / 2T), & 0 < t < 2T \\ 0, & \text{otherwise} \end{cases}, \tag{9}$$

where $a_o$ is the pulse amplitude and $\omega_0 / 2\pi = (f_1 + f_2)/2$ is the carrier frequency. The spectrum associated with the input pulse is shown by the dashed curve in Fig. 2 and has no significant spectral components at the gain lines.

For the parameters of the experiment we find that the first order correction in Eq. (7), i.e. the term proportional to $\partial / \partial t$, is of order $\kappa_1 L / T \approx 10^{-2}$ while the second order correction is smaller than this by $10^{-1}$. Hence, the expansion performed to obtain Eq. (7) is valid.

The differential gain effect, which is misinterpreted as superluminal propagation, requires that $\kappa_1 < 0$. Using Eq. (8) we find that $\kappa_1$ is approximately given by

$$c\kappa_1 \cong -8\pi \frac{(f_1 + f_2)}{(f_2 - f_1)^2} M ,$$

where $\left| f - f_{1,2} \right| >> \gamma$. In this case it is clear that a gain medium $(M > 0)$ is required for $\kappa_1$ to be negative. Note that in the absence of gain $(M < 0)$, $\kappa_1$ can also be negative provided $\left| f - f_{1,2} \right| << \gamma$. In this case differential absorption occurs in which the back of the pulse is absorbed more than the front. This effect has also been presented as superluminal propagation [7].

The validity of Eq. (7) was verified by numerically solving the envelope equation to higher order. Figure 3 compares the solution given by Eq. (7) at the exit of the gain medium (dotted curves) with the vacuum solution $\left| A(0, t - L/c) \right|$ (solid curves). Panel (a) shows the entire pulse profile. Consistent with the experimental measurements, the leading edge is shifted forward in time relative to the vacuum solution by 62 nsec. Panel (b) shows three curves: the solid curve denotes the vacuum solution, the dotted curve shows the result obtained from Eq. (7), and the dashed curve shows the result obtained from Eq. (5). The dotted curve shows that the front of the pulse propagates with velocity $c$; the propagation is not superluminal. The unphysical solution, given by the dashed curve, shows the front of the pulse propagating at superluminal velocities. Panel (c) is an expanded view near the peak of the pulse showing that the front is amplified more than the back.

In conclusion, we find that to properly describe pulse propagation, a consistent ordering of the approximations is necessary. In addition, the distortion of the pulse form that the authors of Ref. 1 misinterpret as a newly observed mechanism for superluminal propagation is actually due to differential gain. That is, the modification of the pulse shape is due to the addition of new photons to the front of the pulse. This phenomenon should not be viewed as superluminal light propagation.

**Acknowledgements**

This work was supported by the Office of Naval Research and the Department of Energy.

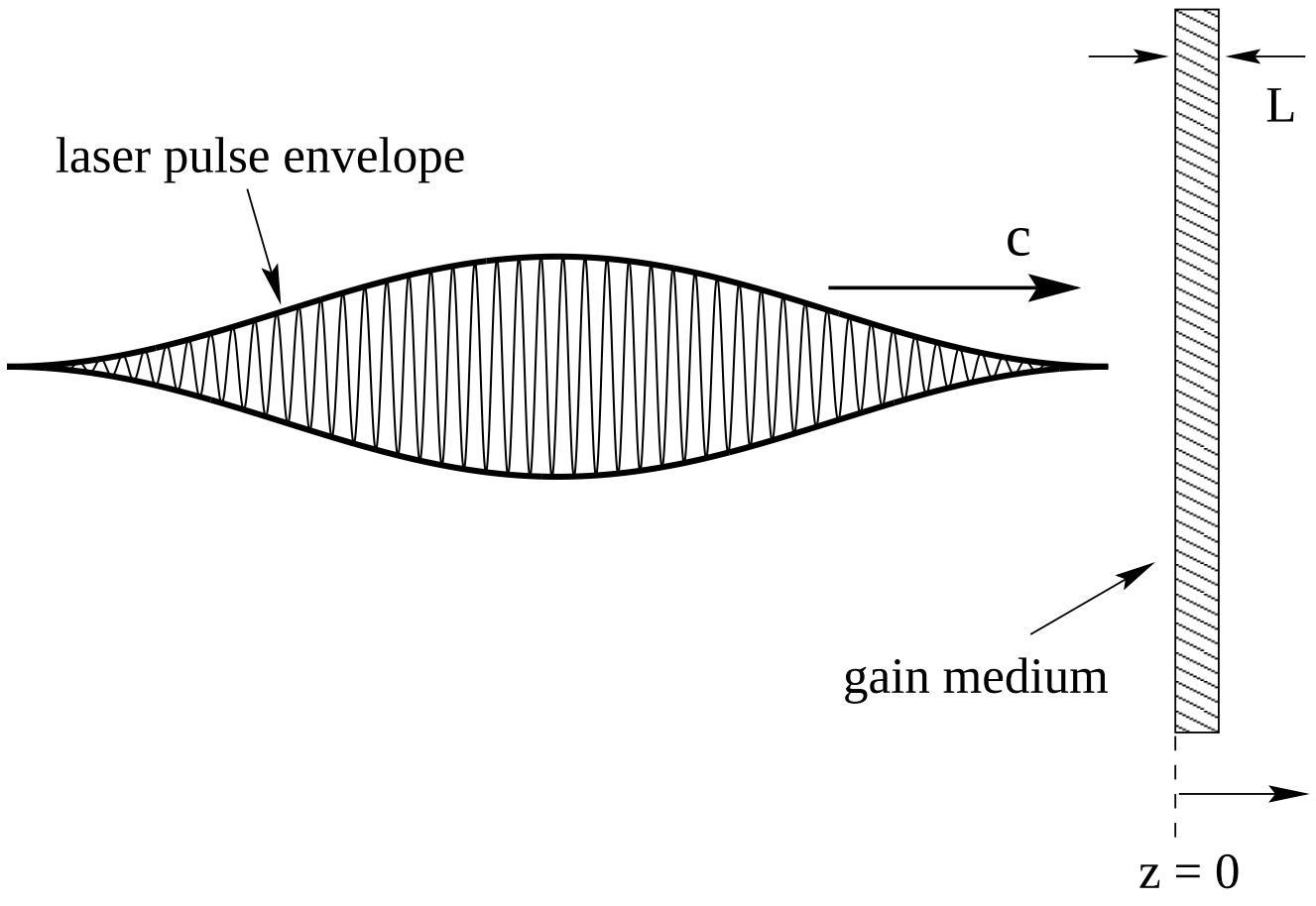

Figure 1. Schematic showing a long laser pulse entering a gain medium.

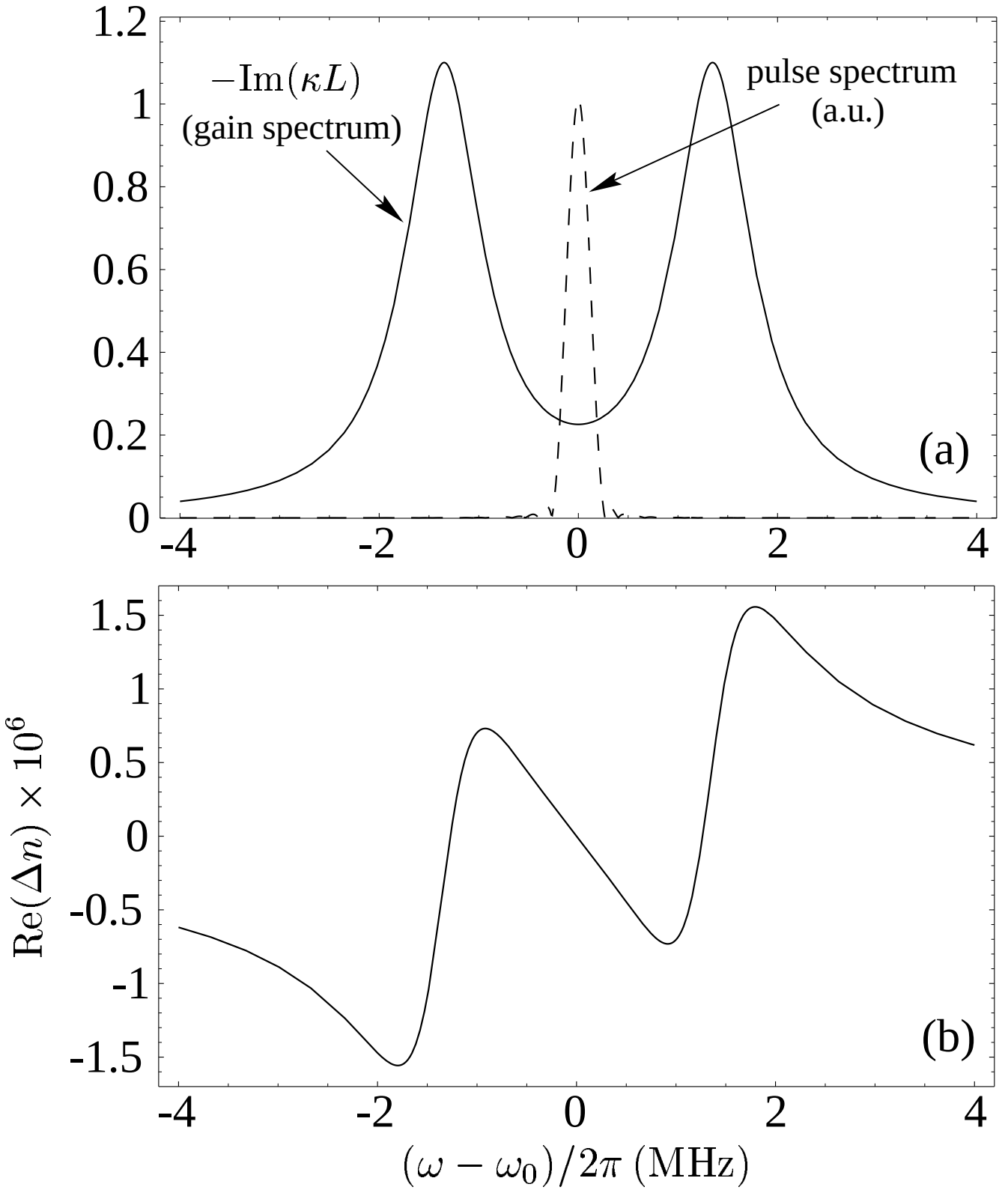

Figure 2. Gain spectrum (solid curve) obtained using the susceptibility in Eq. (8) for the parameters $M = 0.18$ Hz, $f_1 = 3.5\times10^{14}$ Hz, $f_2 = f_1 + 2.7$ MHz, and $\gamma = 0.46$ MHz. The dashed curve shows the spectrum associated with the pulse envelope of Eq. (9) with $T = 3.7$ μsec.

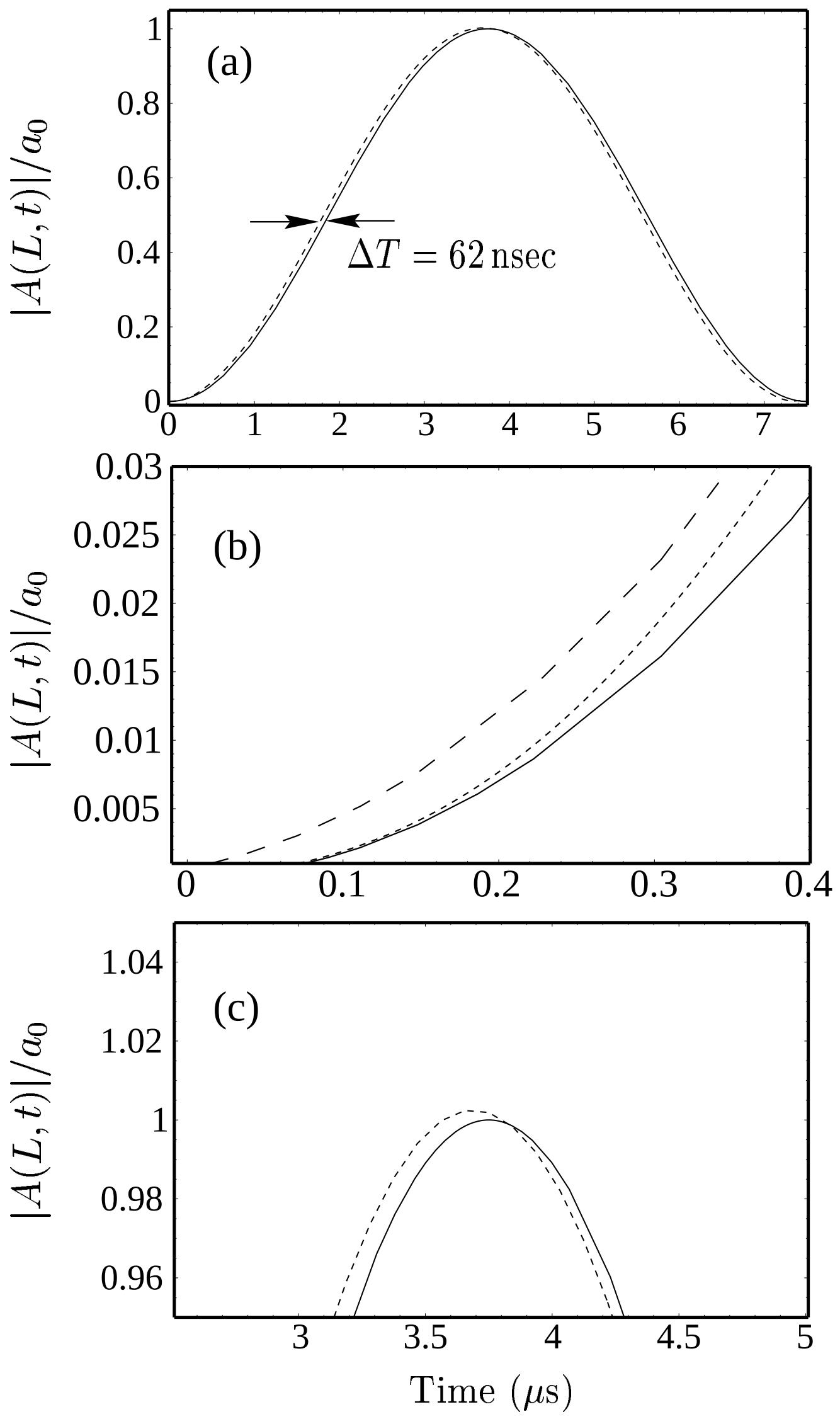

Figure 3: Dotted curves show the pulse envelope amplitude $|A(L,t)|$ at $z = L$ obtained from Eq. (7). Solid curves denote a pulse that has traveled a distance L through vacuum. The dashed curve in panel (b) is the unphysical solution obtained from Eq. (5) showing superluminal propagation. Panels (b) and (c) are expanded views of the front and peak of the pulse, respectively. The parameters for this figure are the same as in Fig. 2.